# Discovering link communities in complex networks by exploiting link dynamics

**Dongxiao He[1], Dayou Liu[1], Weixiongzhang[2,3], Di Jin[4], and Bo Yang[1]**
1 College of Computer Science and Technology, Jilin University, Changchun 130012, People's Republic of China
2 School of Computer Science, Fudan University, Shanghai 200433, People's Republic of China
3 Department of Computer Science and Engineering, Washington University in St Louis, St Louis, MO 63130, USA
4 School of Computer Science and Technology, Tianjin University, Tianjin 300072, People's Republic of China
E-mail: hedongxiaojlu@gmail.com, liudy@jlu.edu.cn, weixiong.zhang@wustl.edu, jindi.jlu@gmail.com and ybo@jlu.edu.cn

**Abstract.** Discovery of communities in complex networks is a fundamental data analysis problem with applications in various domains. Most of the existing approaches have focused on discovering communities of nodes, while recent studies have shown great advantages and utilities of the knowledge of communities of links in networks. From this new perspective, we propose a link dynamics based algorithm, called UELC, for identifying link communities of networks. In UELC, the stochastic process of a link-node-link random walk is employed to unfold an embedded bipartition structure of links in a network. The local mixing properties of the Markov chain underlying the random walk are then utilized to extract two emerged link communities. Further, the random walk and the bipartitioning processes are wrapped in an iterative subdivision strategy to recursively identify link partitions that segregate the network links into multiple subdivisions. We evaluate the performance of the new method on synthetic benchmarks and demonstrate its utility on real-world networks. Our experimental results show that our method is highly effective for discovering link communities in complex networks. As a comparison, we also extend UELC to extracting communities of node, and show that it is effective for node community identification.

# 1. Introduction

Most complex systems in real world, such as people and their complex relationships in a society, molecular elements and their interaction within a cell, and the internet, exist in various forms of networks of nodes connected with edges. One of the main problems in study of complex networks is identification of community structures [1], which has drawn a great deal of interest recently. Although no single definition has been agreed upon, a community within a network is usually considered as a group of nodes that are densely connected with respect to the rest of the network. In the past several years, many approaches have been proposed to uncover community structure in networks, as comprehensively surveyed by Fortunato [2].

Although previous researches for community identification mainly focused on identifying communities of nodes, recent studies began to consider identification of *communities of links* [3, 4]. Link communities are often more intuitive and informative than node communities in practice, for at least two reasons. First, a link usually has a unique identity, i.e., it typically represents a unique relationship between a pair of nodes that the link connects. In contrast, a node may carry multiple identities, i.e., it often plays multiple roles. In a social network, for example, individuals may belong to multiple communities, such as families, friends, and co-workers, while the link between two individuals typically exists for a particular reason which may represent, for instance, a family tie, friendship, or professional relationship. Second, the links connected to a node may carry distinct identities and thus belong to different link communities; consequently, the node can be naturally assigned to multiple communities of nodes when link communities can be identified. Accordingly, overlapping communities of nodes, which form another active research topic of its own [5], could be detected as a byproduct of link communities.

A number of approaches to the identification of link communities have been proposed. Ahn et al. [3] used a hierarchical clustering on a similarity between links to build a dendrogram of links to form a hierarchy of link structure. In order to derive the most relevant communities, they introduced a density function to determine the best level to cut the tree of links. Evans et al. [4, 6] transformed a network of nodes into a line graph using a random walk, and then detected link communities by applying an existing algorithm for node partition on this line graph. Kim et al. [7] extended the map equation method [8], which was originally developed for node communities, to link community identification by assigning the communities to links instead of nodes, modifying the encoding rule for the random walk to represent this change in the community structure, and proposing the corresponding map equation for the link community. Pan et al. [10] detected link communities by local optimization. They first selected some links as seeds for a network, and then expanded these selected seeds by greedily maximizing a local community fitness function, to attain all link communities. Finally, an important algorithm, which is perhaps representative of the state-of-the-art in this area, is due to Newman et al. [9]. This is a principled statistical approach using generative network models and implemented in a closed-form expectation-maximization algorithm. This algorithm will be used in our current study for comparison.

Despite the previous works on link community identification, properties of link communities remain to be fully explored and exploited. In this paper, we consider the relationship between the property of Markov dynamics and community structures, so as to reveal intrinsic partitions of links in a network. First, we introduce a stochastic process of link-node-link random walk, and utilize its aggregated transition probability distribution to unfold a bipartition structure of links for a network. We then analyze local mixing properties of the Markov chain of this stochastic process and design a suitable method to extract two emerging link communities. Integration of the two components forms a bipartition scheme; an iterative application of this bipartition scheme constitutes a novel method for link community identification. For comparison, we also extend this method to finding node communities.

# 2. The UELC algorithm

*2.1. Overview of the UELC*

Let $N = (V, E)$ be an undirected and unweighted network, where $V$ is the set of $n$ nodes (or vertices) and $E$ is the set of $m$ edges (or links), and let $A$ denote the adjacency matrix of $N$. Instead of assuming that a community is a set of nodes densely connected to one another, here we consider a community to be a set of closely interrelated links, which is called a link community. The identification of link communities is an extension to the conventional concept of community identification, whose objective is to find a division of network links into communities.

A straightforward approach to this problem, which is the one adopted in our UELC algorithm, is the repeated division of a network. We first divide a network into two link communities, and then recursively subdivide the two parts. In dividing a subnetwork, we isolate it from the rest of the network and perform a 'nested' UELC on it, resulting in a link partition of the subnetwork with two smaller link communities. Subsequently, we decide whether or not to accept this bipartition by a special method based on link density. A key to UELC is to design a suitable method that could precisely partition the links of a network into two parts with a high link density. Here, we employ a Markov dynamics based method, which contains two phases. In the first or ULC phase, we utilize a stochastic process of link-node-link random walk to reveal a bipartition structure of links for a network. In the second or ELC phase, we exploit the local mixing properties of the underlying Markov chain underlying the random walk to extract the two emerging link communities. We summarize the UELC algorithm in the form of recursion, which is listed in Appendix A.

*2.2. Bipartition a network*

We now introduce a Markov dynamics based method to bipartition the links of a network. The method consists of two phases, ULC and ELC.

*2.2.1. Unfold a bipartite structure*

Here, we consider a stochastic process of a link-node-link random walk on a network, and utilize its aggregated link transition probability distributions to reveal a bipartite structure of the links in the network.

Given a network $N = (V, E)$, consider a stochastic process defined on $N$, in which an agent freely walks from one *link* to another through the nodes they share. When the agent arrives at a link, it will randomly select one of its two endpoints, and then randomly choose one of the links connected to this node and move there. This is one step of a link-node-link random walk, which is considered as one step in the stochastic process.

Assume that $X = \{X_t, t \geq 0\}$ represents the agent positions of edges, and $P\{X_t = e_{ij}, e_{ij} \in E\}$ denotes the probability that the agent arrives at link $e_{ij}$ after exactly $t$ steps. For $t > 0$ we have $P\{X_t \mid X_0, X_1, \ldots, X_{t-1}\} = P\{X_t \mid X_{t-1}\}$, i.e., the walking process is Markovian so that the next state of the agent is only dependent of its current state. Furthermore, $X_t$ is homogeneous because $P\{X_t = e_{ij} \mid X_{t-1} = e_{pq}\} = p(e_{pq}, e_{ij})$, where $p(e_{pq}, e_{ij})$ is the transition probability from $e_{pq}$ to $e_{ij}$ by one step of a link-node-link walking and can be defined by the adjacency matrix of $N$, $A = (a_{st})_{n \times n}$, as follows,

$$p(e_{pq}, e_{ij}) = \begin{cases} \dfrac{1}{2} \cdot \dfrac{a_{ij}}{\sum_r a_{ir}} + \dfrac{1}{2} \cdot \dfrac{a_{ij}}{\sum_r a_{jr}} & \text{if } \{i,j\} \cap \{p,q\} = \{i,j\} \\ \dfrac{1}{2} \cdot \dfrac{a_{ij}}{\sum_r a_{ir}} & \text{if } \{i,j\} \cap \{p,q\} = \{i\} \\ \dfrac{1}{2} \cdot \dfrac{a_{ij}}{\sum_r a_{jr}} & \text{if } \{i,j\} \cap \{p,q\} = \{j\} \\ 0 & \text{otherwise} \end{cases} \quad (1)$$

where 1/2 is the probability that the agent randomly selects one of the two endpoints of $e_{pq}$ and $a_{ij}/\sum_r a_{ir}$ (or

$a_{ij} / \sum_r a_{jr}$) is the probability that the agent randomly chooses one of the links connecting with this node. As we only consider undirected and unweighted networks, the values of the matrix elements $a_{st}$ are equal to zero or one, and link $e_{ij}$ and link $e_{ji}$ are regarded as identical.

Consider the above Markov process again. Let $\alpha^l(e_{ij})$ be the probability that the agent starts from a source position and eventually arrives at an destination link $e_{ij}$ with $l$ steps. The value of $\alpha^l(e_{ij})$ can be estimated iteratively by

$$\alpha^l(e_{ij}) = \sum_{r \neq j} \alpha^{l-1}(e_{ri}) \cdot p(e_{ri}, e_{ij}) + \sum_{s \neq i} \alpha^{l-1}(e_{sj}) \cdot p(e_{sj}, e_{ij}) + \alpha^{l-1}(e_{ij}) \cdot p(e_{ij}, e_{ij}) \quad (2)$$

The first term of the right side of (2) is the probability that the agent arrives at $e_{ij}$ from its neighbor links sharing node $i$, the second term is the probability that the agent arrives at $e_{ij}$ from its neighbor links sharing node $j$, and the third is the probability that the agent arrives at $e_{ij}$ from itself. Assume that the source or starting position is link $e$. When step number $l$ is 0, $\alpha^0(e_{ij})$ is equal to 1 if $e_{ij} = e$, or 0, otherwise. Here, $\boldsymbol{\alpha}^l$ is called the $l$ step transition probability distribution (vector). Note that the sum of the probabilities that the agent arrives at all the links from a source position will be 1, i.e. $\sum_{e_{ij} \in E} \alpha^l(e_{ij}) = 1$.

Let a network possess an obvious bipartition structure that consists of two separable link communities. As each link community has a higher edge density than the entire network, an agent that starts from a source position should have more link-node-link paths to choose from the community where the starting edge belongs to in order to reach the edges in its own link community with $l$ steps, when the value of $l$ is adequate. On the contrary, the agent should have a lower probability to arrive at the edges outside its affiliated link community. In other words, it will be difficult for the agent to escape from its own link community by passing those "bottleneck" link-node-link paths through the border nodes to arrive at the other part of the network. Thus, in general, when the step number $l$ is suitable, vector $\boldsymbol{\alpha}^l$ should meet the following condition

$$\underset{e_{ij} \in C_{in}}{\forall} \underset{e_{pq} \in C_{out}}{\forall} : \alpha^l(e_{ij}) > \alpha^l(e_{pq}) \quad (3)$$

where $\boldsymbol{C_{in}}$ denotes the link community where the source edge belongs to, and $\boldsymbol{C_{out}}$ denotes the rest of the network edges.

We now proivde an example, shown in Figure 1, to illustrate the aforementioned mechanism. This network consists of two evident link communities: the red and the blue communities. The density of each link community is evidently higher than that of the entire network, and the link-node-link paths that connect these two communities can be regarded as "bottlenecks" that prevent the agent from escaping its current link community. Consider a vast collection of random walking agents, who start from the same link (e.g., $e_{14}$). The agents will in general follow different link-node-link paths. Over the course of such random walks, the majority of the crowd will tend to stay in the same (red) community as it is difficult for the agents to pass through the bottlenecks. This phenomenon is analogous to pouring water into a river channel whose topology is defined by Figure 1. Within suitable time, the current in red region will be greater than that of blue region, if we pour water into the red area.

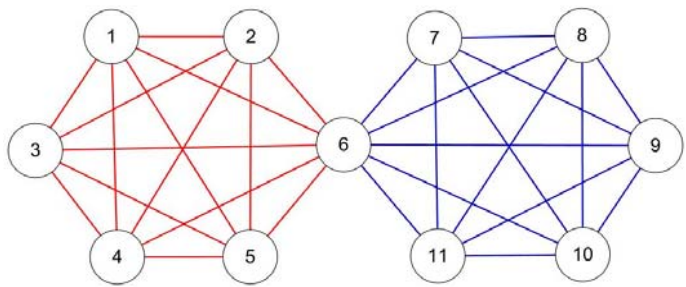

**Figure 1.** A sample network with two evident link communities.

Our method, which was called ULC, follows the above idea to unfold a bipartition link structure of a network. Its takes the following main step: 1) randomly select a link $e$ and take it as the source position for the agent and 2) calculate the corresponding $l$ step transition probability vector $\alpha^l$ according to (2). Based on (3), when the step number $l$ is adequate, the majority of the links within the link community containing the source position should have higher probabilities to be reached than the links in the rest of the network, so that a bipartition structure of the network will thus emerge and be detected. In the next section, we will introduce an effective strategy for extracting the two emerged link communities and a method for determining an adequate step number $l$.

We now further depict the idea of ULC with a real-world, simple network, i.e., the well-known Zachary's "karate club" network. This is a network of friendships between 34 members of a university sports club, which was constructed according to the observational study of Zachary. Nodes denote members of the club, and links connect any pair of members which consistently were observed to interact outside the normal activities of the club. During the course of Zachary's study, a dispute between the club's administrator and its karate instructor arose. Some members support the opinion of the administrator, while the others support the instructor's. Finally, the club split into two factions, centered on the administrator and instructor respectively. Figure 2(a) shows this network with its actual splitting result of nodes [15] as well as a link partition obtained by Newman et al. [9]. In this figure, node communities are distinguished by different shapes of nodes, and link communities are shown by different colors on links. As shown, the Newman's method classified all links connecting circles or squares into the red or blue community, respectively. Meanwhile, it classified the links that connect nodes of different shapes into reasonable communities. To be specific, for links whose circle endpoint has a larger proportion of neighbors with the same shape as it (this endpoint can be considered having a more clear standpoint about supporting the administrator or the instructor), the Newman's method classified them into the red link community; for links whose square endpoint has a more clear standpoint, the method classified them into the blue community. Therefore, we take the Newman's result as a presumable ground truth of link communities of the network. Figure 2(b) shows the calculated $l$ step transition probability distribution $\alpha^l$, starting from a randomly selected red link. As shown, the link probabilities for the red link community are greater than that for the blue community. It is evident that $\alpha^l$ meets condition (3) and can be used to reveal a clear bipartition link structure of this network; there is a clear separation between the probabilities for the two communities.

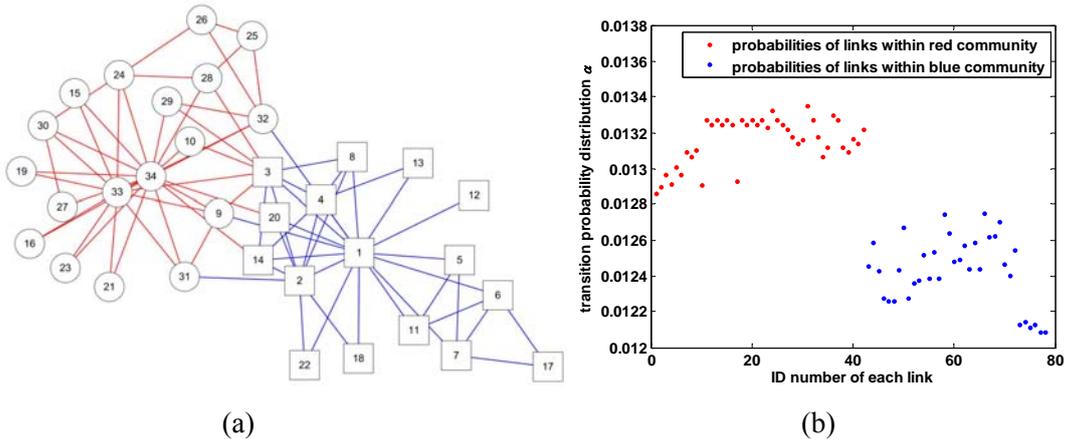

(a)          (b)

**Figure 2.** An illustration of the ULC method for revealing a bipartition link structure of the well-known Zachary's "karate club" network. **(a)** The network itself, drawn by Cytoscape [11] with the force-directed layout [12]. It contains two actual node communities, marked by circle and square nodes, respectively. It also includes two presumably true link communities, colored in red and blue, based on the analysis of a method by Newman et al. [9]. **(b)** The generated $l$ step transition probability vector $\alpha^l$ according to (2), where the step number $l$ is set to 16, which is a lower bound of this parameter for this network.

**Proposition 1.** The time complexity of ULC is $O(lmk)$, where $l$ is the step number, $m$ is the number of links in network $N$, and $k$ is the maximum degree of all nodes of $N$.

***Proof.*** For $p$ in (1), there are $\sum_{i=1}^{n}(k_i(k_i-1)/2)+m$ nonzero elements to be calculated, where $k_i$ is the degree of node $i$. As $k_i = \sum_r a_{ir}$ for all nodes can be calculated in advance in time $O(m)$, each element of $p$ can be computed in a constant time. Thus, the time to calculate $p$ according to (1) is $O(\sum_{i=1}^{n} k_i^2)$, which can be also given by $O(mk)$ as $\sum_{i=1}^{n} k_i^2 \leq k\sum_{i=1}^{n} k_i = mk$. Computation of vector $\alpha^l$ by (2) can be done in $l$ iterations. To compute any element $\alpha^t(e_{ij})$ for each iteration $t$, we need to consider all neighbors of link $e_{ij}$. Thus, computation of vector $\alpha^t$ requires to traverse all elements of $p$ twice, which takes $O(mk)$ time. Therefore, the total time to compute vector $\alpha^l$ is $O(lmk)$, which is also the time complexity of ULC. QED

### 2.2.2. Extract two emerged communities

In order to extract emerged link communities, we first transform the above stochastic process of link-node-link random walk on the original network $N$ to an equivalent stochastic process of node-node random walk on its weighted line graph $L$ to analyze the local mixing properties of the resulting Markov chain. Then, we make use of the local mixing properties of the Markov chain to estimate a theoretical lower bound of step number $l$, which is subsequently used to derive the $l$ step transition probability distribution $\alpha^l$ for identifying link communities.

Given a network $N$, its line graph $L$ is such a network that each node of $L$ represents an link of $N$ and two nodes of $L$ are adjacent when their corresponding links share a common endpoint in $N$. Different from traditional unweighted line graph, here we take $p_{m \times m}$ calculated by (1) as the adjacency matrix of $L$. A simple example is shown in Figure 3. Consider now a new stochastic process defined on $L$, in which an agent walks from one node to another along the links between them. When the agent arrives at a node, it will randomly select one of its neighbors and move there, and the process repeats to form a conventional node-node random walk.

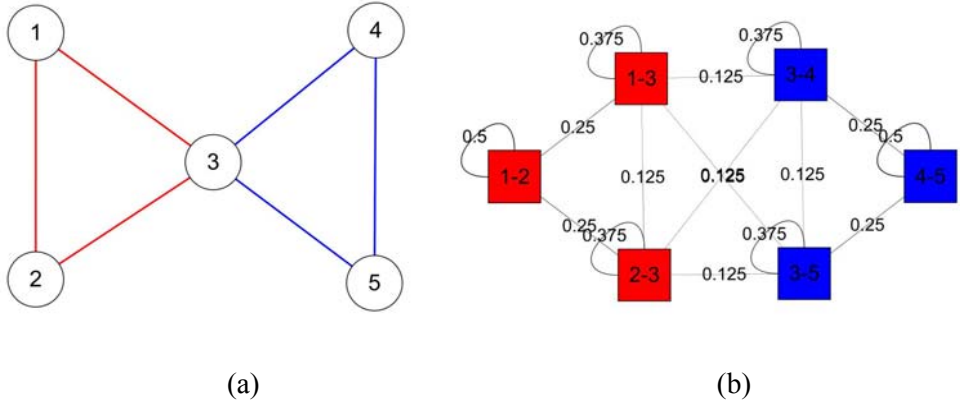

(a)        (b)

**Figure 3.** An example of transforming a toy network into its weighted line graph. **(a)** A network consisting of two link communities in red and blue. **(b)** The corresponding line graph with edge weights according to (1). The width of an edge is depicted by its weight, and a node color corresponds to the color of its associated link in the original network.

Consider two nodes $u$ and $v$ in line graph $L$, which respectively correspond to link $e_{pq}$ and link $e_{ij}$ in the original network $N$. Let $q(u, v)$ be the transition probability from node $u$ to node $v$, which can be computed as

$$q(u,v) = \frac{p(u,v)}{\sum_r p(u,r)} \qquad (4)$$

It is evident that $q(u, v)$ is equal to $p(e_{pq}, e_{ij})$ because $\sum_r p(u,r) = 1$. Let $\beta^l(v)$ denote the probability that the agent starts from a source node and eventually arrives at an destination node $v$ with $l$ steps, we have:

$$\beta^l(v) = \sum_{r \in Neb(v)} \beta^{l-1}(r) \cdot q(r,v) \tag{5}$$

where **Neb(v)** is the set of neighboring nodes of $v$. Because $v = e_{ij}$, there is **Neb(v)** = $\{e_{xi} \mid x \neq j\} \cup \{e_{yj} \mid y \neq i\} \cup \{e_{ij}\}$. Also, $q(u, v)$ is equal to $p(e_{pq}, e_{ij})$ for $u = e_{pq}$ and $v = e_{ij}$. Thus, $\beta^l(v)$ is equivalent to $\alpha^l(e_{ij})$ in (2). Then, our link-node-link random walk on network **N** can be regarded as a conventional node-node random walk on its weighted line graph **L**. Thus, we can analyze the local mixing properties of the Markov chain of link-node-link random walk with the aid of the existing theories on conventional node-node random walk, so as to design a suitable method to extract emerged link communities.

According to [13], the local mixing properties of the Markov chain of a node-node random walk on a network are associated with the community structure of the network. The dynamics of the Markov chain of a node-node random walk on a network such as **L** can be described as follows: before the Markov chain reaches its global mixing state $s_1$, it should go through $m-1$ local mixing states first along the time dimension, denoted as $s_m, \ldots, s_3, s_2$ ($m$ is the number of nodes in **L**). Here, the $i$-th local mixing state $s_i$ corresponds to $i$ locally mixed communities of a network.

Further, according to the theory of [13], all the local mixing times of the Markov chain can be estimated by using the spectrum of its Markov generator (normalized graph Laplacians) $M = I - Q$, where $I$ is the identity matrix and $Q$ is the transition probability matrix $q_{m \times m}$ calculated according to (4). For undirected network, $M$ is positive semi-definite and has $m$ non-negative real-valued eigenvalues ($0 = \lambda_1 \leq \lambda_2 \leq \ldots \leq \lambda_m \leq 2$). Let $T_i^{ent}$ and $T_i^{ach}$ be the entering time and achieving time of the $i$-th local mixing state. It has been shown in [13] that

$$T_i^{ach} = \frac{1}{\lambda_i}(1 + o(1)) \tag{6}$$

Further, we can also use the achieving time of the $(i+1)$-th local mixing state to estimate the entering time of the $i$-th local mixing state. That is $T_i^{ent} = T_{i+1}^{ach} = 1/\lambda_{i+1}$.

In particular, $T_2^{ach} = 1/\lambda_2$ is the time that the Markov chain achieves the second local mixing state and begins to enter the global mixing state, and $T_1^{ach} = 1/\lambda_1 = \infty$ is the time that the Markov chain arrives at the global mixing state. In other words, each of two communities in a network has locally mixed at time $T_2^{ach}$. While these two communities can never be globally mixed, they can gradually approach each other from time $T_2^{ach}$ to $\infty$. For probability vector $\beta^l$, starting at time $T_2^{ach}$, the associated probability at a node in the same or different community as the source node should be greater or less than the probability at the global mixing state, respectively. Thus, if we take $T_2^{ach}$ as a lower bound of the step number $l$, we can use the global mixing state to separate the two locally mixed communities. According to [13], for an undirected network, the global mixing state of its Markov chain can be computed as

$$\beta^{\infty}(v) = \frac{d_v}{\sum_r d_r} \tag{7}$$

where $d_r$ is the weighted degree of node $r$. For any node $r$ in line graph **L**, $d_r = \sum_s p(r,s) = 1$. Thus, we have $\beta^{\infty}(v) = 1/m$ for any node $v$. Furthermore, for the original network **N**, $\alpha^{\infty}(e_{ij}) = \beta^{\infty}(v) = 1/m$ for any link $e_{ij}$.

Based on the above idea, we can devise an approach, namely ELC, to extract two emerged link communities in network **N**. This method takes the stable probability $\varepsilon = 1/m$ as a threshold to divide the $l$ step transition probability vector $\alpha^l$ into two parts; those links whose associated probabilities are greater than $\varepsilon$ belong to a community, and the links whose associated probabilities are less than $\varepsilon$ are classified to the other community. It is obvious that the time complexity of ELC is $O(m)$.

Again we use the example in Figure 2 to illustrate ELC for extracting two emerged link communities. Figure 4(a) shows the weighted line graph of the karate club network. The network has two communities of nodes, corresponding to the two communities of links in its original network in Figure 2(a). The black line in the middle of Figure 4(b) corresponds to the stable probability $\varepsilon = 1/m$, which is adopted as a threshold to bipartition the $l$ step transition probability vector $\alpha^l$ from Figure 2(b). As shown, when $l$ is set to this theoretical lower bound of $\lceil 1/\lambda_2 \rceil = 16$, the two communities have been locally mixed, respectively, with the aid of the global mixing state of the Markov chain.

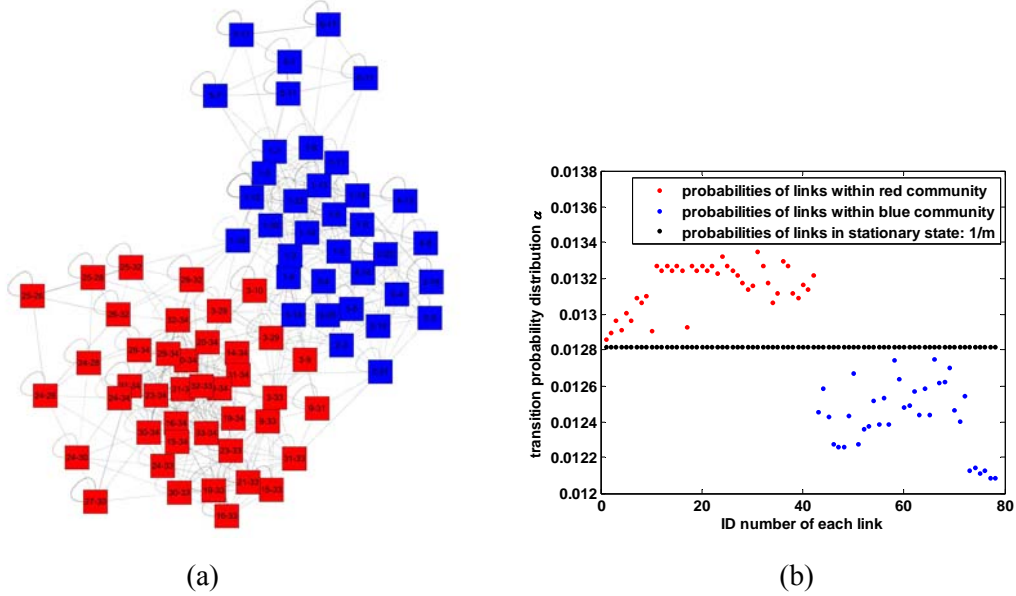

(a)          (b)

**Figure 4.** An example of the ELC method to extract two emerged link communities in the network of Figure 2(a). **(a)** The corresponding line graph of Figure 2(a). The colors of nodes correspond to the colors of their associated links in the original network. **(b)** The global mixing state of the Markov chain is used to separate the two locally mixed communities. The black line denotes the stable probability $\varepsilon = 1/m$, which is used as a threshold to bipartition the $l$ step transition probability vector $\alpha^l$ in Figure 2(b). Note that step number $l$ is set to 16 as $1/\lambda_2$ is 15.1203 for the weighted line graph.

We also need to evaluate the theoretical lower bound ($1/\lambda_2$) for step number $l$. By using one of the fastest algorithms, namely Lanczos method [14], the time to calculate the second smallest eigenvalue ($\lambda_2$) of $M$ is $O(m'/(\lambda_3-\lambda_2))$, where $m' = \sum_{i=1}^{n}(k_i(k_i-1)/2) + m$ is the number of edges in line graph $L$. This time complexity depends on not only the scale of $L$ but also the difference between the second and the third eigenvalues, and can be simplified to $O(km/(\lambda_3-\lambda_2))$. Our experimental analysis indicates that the theoretical lower bound ($1/\lambda_2$) for step number $l$ is usually very small (often smaller than 100), and seems to be independent to the scale of networks. For example, the values of $1/\lambda_2$ for line graphs of networks of karate club [15], *Les Misérables* [16] and word association [17] are 15.1203, 22.6927 and 12.7181, respectively. Thus, in order to avoid to compute eigenvalue $\lambda_2$ which can be time consuming, we generally set an upper bound on step number $l$ to a large constant (e.g., 100), as our method is insensitive to the value of $l$ when it is greater than $1/\lambda_2$.

*2.3. Terminal condition for the subdivision process*
A key element of algorithm UELC is the termination condition for the repeative process of subdividing the links of a network $N$, so as to obtain a superior link community structure. Modularity $Q$ has been widely used for similar purposes in node communities [18, 19], and it can also be used to evaluate the quality of link communities in the weighted line graph of an original network [4]. But this measure, as well as its multiresolution versions [20], suffers from resolution limits [21, 22] and may exhibit extreme degeneracies [23]. Furthermore, modularity $Q$ is not *a priori* consistent with the way we find link communities. Thus, a new, specialized stopping criterion is in demand here.

Recall the main idea of our Markov bipartition method. When a network possess two clear link communities, each of which has a higher edge density compared to the entire network, a random walk agent that starts from a source position should have a higher probability to stay within the community where the source position belongs to than to move outside of the community. This property is exploited in our Markov bipartition method. On the contrary, should this property fail to hold on a network of interest, our method would not work, and the network could be regarded as non-partitionable. Based on this observation, we set the termination condition of the subdivision process to the condition of detecting one part of the local bipartition result has a lower density, to be defined precisely below, than that of the current network (or subnetwork). Note that community densities have been regarded as a good quality measure for community identification, which do not suffer from resolution limit [3].

We define the density of a link community (or network) in the spirit of the partition density proposed by Ahn et al [3]. Let a network $S$ have $n_s$ nodes and $m_s$ edges. Then, its link density can be computed by

$$D_s = \frac{m_s - (n_s - 1)}{n_s(n_s - 1)/2 - (n_s - 1)} \tag{8}$$

The link density $D_s$ is $m_s$ normalized by the minimum and maximum numbers of links among $n_s$ connected nodes. Thus, $D_s=1$ when network $S$ is a clique, or $D_s=0$ when $S$ is a tree. In particular, we assume that $D_s=0$ if $n_s=2$ without loss of generality. In essence, $D_s$ measures how "clique-ish" vs. "tree-ish" that network $S$ is.

**Proposition 2.** The time complexity of UELC is $O(Tlmk)$, where $T$ is the number of link communities detected, $l$ is the step number, $m$ is the number of edges in network $N$, and $k$ is the maximum degree of all nodes of $N$.

***Proof.*** From the previous section, the time to bipartition a link network $N$ is $O(lmk+m)$. The time to decide whether to accept a bipartition result is a constant. Because $T$ link communities are detected, there should be $(2T-1)$ subdivision operations in UELC. Also, the size of a subnetwork to be considered decreases in the subdivision process. Thus, the time complexity of UELC should not be greater than $O((2T-1)\cdot(lmk+m+1))$, which gives rise to $O(Tlmk)$.

## 3. Experimental analyses

In order to evaluate the performance of UELC, we tested it on benchmark synthetic networks and some widely used real-world networks. The synthetic networks allow us to test UELC's ability to detect known communities under controlled conditions, while the real networks allow us to observe its performance under practical conditions.

All experiments were done on a single Dell Server with Intel(R) Xeon(R) CPU 5130 with a 2.00GHz processor and 4 Gbytes of main memory. The source code of the algorithms used here can all be obtained from the authors and is also available online [24].

*3.1. Synthetic networks*

The synthetic networks we used for the tests were generated based on the link community model [Ball-Karrer-Newman (BKN) model], proposed in [9]. Also, the expectation-maximization (EM) algorithm [9] based on the same stochastic model that was used to generate the synthetic networks was selected here to be compared with our algorithm UELC. Furthermore, we employed two accuracy measures used in [9], i.e., "Fraction of Vertices Classified Correctly (FVCC)" and "Jaccard index", to compare the quality performance of the two algorithms.

Following the experiment design of [9], we set the parameters for the BKN model as follows. The networks have $n = 10000$ nodes each, divided into two overlapping (link) communities. We placed $x$ nodes in the first community only, i.e., these nodes have connections exclusively within the community, $y$ nodes in the second community only, and the remaining $z = n-x-y$ nodes in both communities, with equal numbers of connections to nodes in these two communities on average. We set the expected degree of all nodes to a fixed value $<k>$. We varied the parameters $x$, $y$, $z$, and $<k>$ to generate networks with stark community structures or no structure at all, so as to vary the difficulty of the network instances posed to the algorithms.

We performed three sets of tests. In the first set of experiments, we fixed the size of the overlap between the communities at $z = 500$, divided the remaining nodes evenly (i.e., $x = y = 4750$), and varied the value of $<k>$ from 1 to 15 with an increment of 1. For the second set of tests, we again set the overlap at $z = 500$ but fixed $<k> = 10$ and varied the ratio between $x$ and $y$. Finally, for the third set of tests, we set $<k> = 10$, constrained $x$ and $y$ to be equal, and varied the amount of overlap $z$.

As the EM algorithm requires a *priori* information, i.e., the number of communities, we set the parameter of community number of EM as 2 which is the actual number of communities of the benchmark networks in the following experiments. For the sake of fairness, our UELC provided its bipartition result to compare with the link partition obtained by EM. In Figure 5, we show the fraction of corrected classified nodes by the two algorithms for each of the three sets of experiments. To be considered correctly classified, a node's membership in both communities must be reported correctly by an algorithm. As shown in Figure 5, UELC outperforms EM in terms of FVCC accuracy in the first and third tests. In the second test, UELC has a better performance than EM in 7 out of 10 network configurations. Furthermore, we adopted the Jaccard index to compare the two algorithms' ability for identifying overlapping communities using the same sets of network instances. If $S$ is the set of truly overlapping nodes and $V$ the set of predicted overlapping nodes, then the Jaccard index is $J=|S \cap V|/|S \cup V|$. This index is a standard measure of similarity between sets that rewards accurate identification of the overlap while penalizing both false positives and false negatives. Figure 6 shows the result of comparing the two algorithms using the Jaccard index. As shown, UELC is superior to EM in the first and third sets of experiments; UELC also outperforms EM when the community size is relatively balanced in the second set of experiments. This result is similar to the result in Figure 5, and they both confirm the validity of our new algorithm UELC.

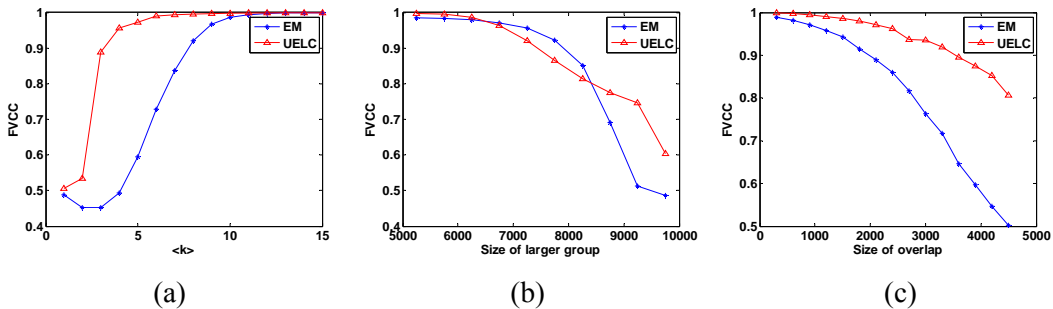

(a)　　　　　　　　　　(b)　　　　　　　　　　(c)

**Figure 5.** Comparison of UELC and EM algorithms in the three sets of synthetic networks, measured by the fraction of vertices classified correctly. Each data point in the figure is an average over 50 network instances. (a) The FVCC accuracy as a function of the expected degree $<k>$ of all nodes. (b) The FVCC accuracy as a function of the size of the larger community. (c) The FVCC accuracy as a function of the amount of overlap between the two communities.

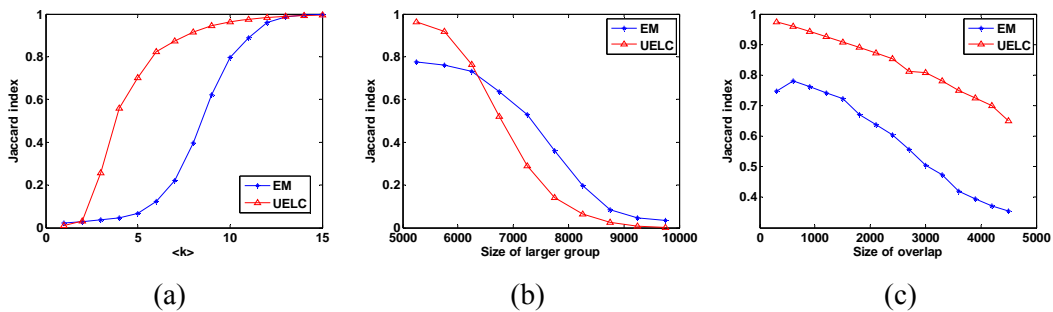

(a)　　　　　　　　　　(b)　　　　　　　　　　(c)

**Figure 6.** Comparison of UELC and EM algorithms in the three sets of synthetic networks measured by the Jaccard index. Each data point in the figure is an average over 50 network instances. (a) The Jaccard index as a function of the expected degree $<k>$. (b) The Jaccard index as a function of the size of the larger community. (c)

The Jaccard index as a function of the amount of overlap.

*3.2. Real-world networks*

We further evaluated our new algorithm UELC on three real-world networks, i.e., Zachary's "karate club", *Les Misérables* and word associations.

*3.2.1. Karate club*

Zachary's "karate club" network [15] which has been used in Sec. 2.2.1 has almost become a bestbed for community detection algorithms. Thus we considered it in our first experiment. The description of this network is in Sec. 2.2.1. Figure 7 shows the network with the instructor and the administrator represented by nodes 1 and 34 respectively, and the actual community structure in terms of nodes, with circles representing members of the administrator's faction and squares representing members of the instructor's faction.

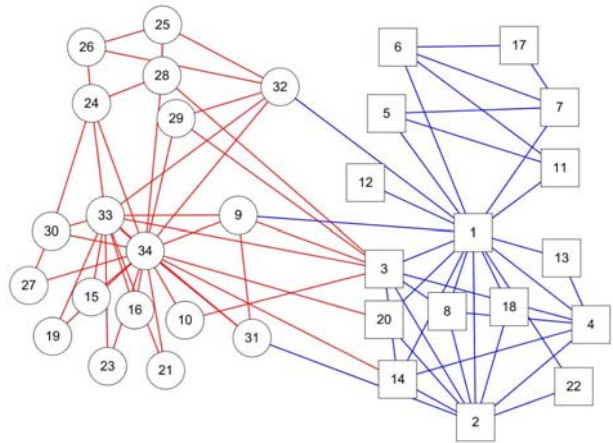

**Figure 7.** The Zachary's "karate club" network. It demonstrates the actual split of nodes which contains two communities marked by circle and square nodes respectively. Also, the bipartition result of links attained by UELC is shown by different colors on links.

We applied our algorithm UELC to this interesting network. The bipartition result is shown in Figure 7 with links colored accordingly. The bipartition result is identical with the link partition obtained by Newman et al. [9]. According to the above analysis in Sec. 2.2.1, this bipartition is reasonable. In addition, the result reveals six overlapping nodes, i.e., No.3, 9, 14, 20, 31 and 32. In fact, the individuals represented by these overlap nodes have friends in both camps. Further, Zachary's original discussion of the split includes some indications that these individuals have had some difficulty deciding which side of the dispute to come down on [9]. Thus, it is reasonable to mark these nodes as overlap nodes.

After the bipartition of "karate club" network, each link community in the above two-way partition is further subdivided until the termination condition is satisfied. The final result from UELC is shown as Figure 8. It is a four-way partition with a higher resolution, in which each link community has a higher link density compared with its parent community in the previous bipartition step. Furthermore, the four-way partition reveals some new, subtle community structure that is missing in the bipartition result. For example, the club's instructor (node 1) and administrator (node 34), who are identified as overlapping nodes, connect to different communities of people and reside at the centers of their respective communities, each of which can be further divided into subgroups.

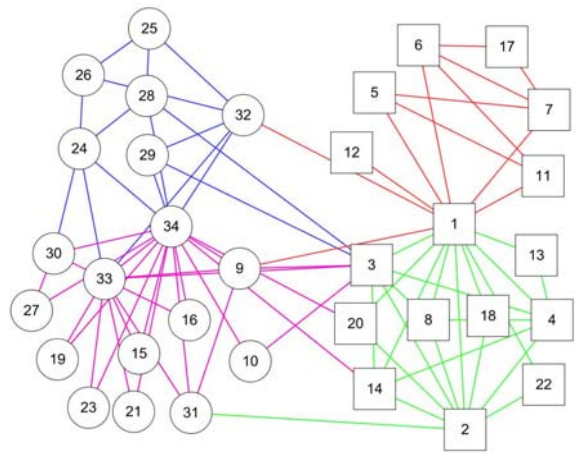

**Figure 8.** The link partition obtained by UELC for the "karate club" network.

*3.2.2. Les Misérables*

In our next experiment, we considered a co-appearance network of 77 characters in the novel *Les Misérables* by Victor Hugo. This network was constructed based on the list of characters' appearance by scene complied by Knuth [16]. In this network, nodes denote characters and a link between two nodes represents co-appearance of the corresponding characters in one or more scenes. Figure 9 shows the partition from UELC, which divides the network into 5 overlapping communities. The resulting partition agrees reasonably well with social divisions and subplots in the plot-line of the novel.

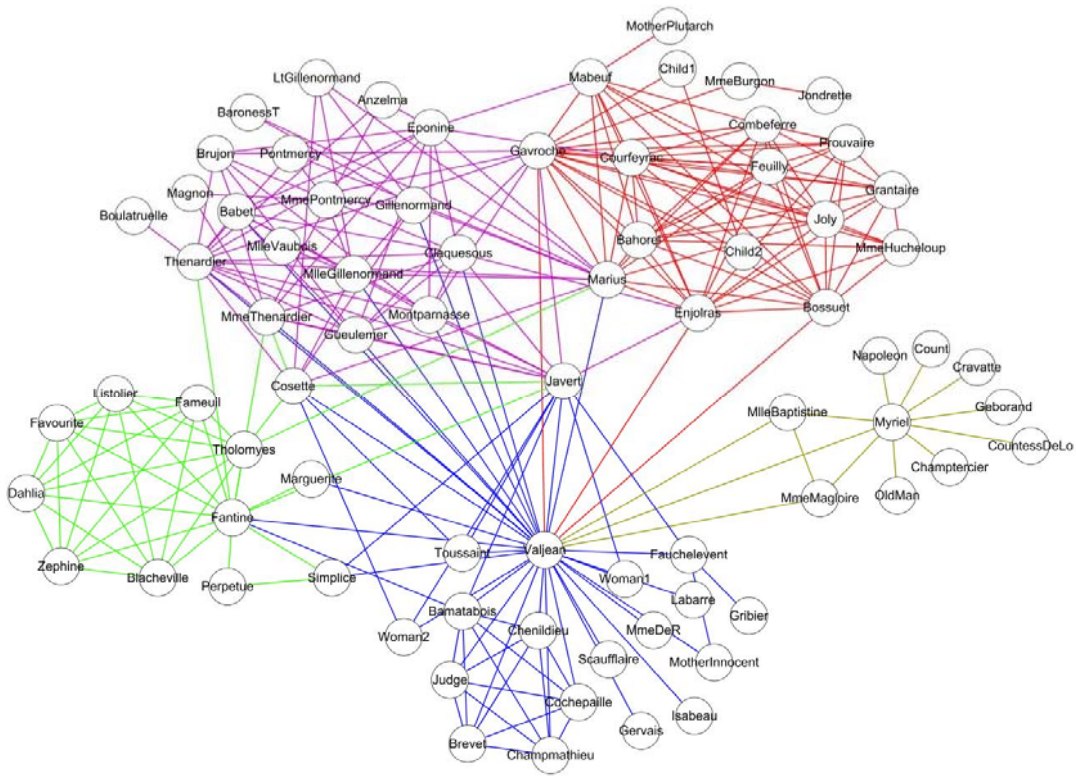

**Figure 9.** The link communities detected by UELC for the network of characters from *Les Misérables*.

Similar to many networks in the real world, this network contains some highly connected nodes, some of which, e.g., Valjean, can connect to most parts of the network. These nodes can cause serious problems for conventional node community schemes because they do not fit adequately to any community. No matter where we place a highly connected node, it is going to have many of its neighbors in other communities. In contrast, link community schemes can provide an elegant solution to this problem because they allow a node to hold multiple positions in different communities. Therefore, we can place the hubs (i.e., the nodes with high degrees) in overlaps between communities. As shown in Figure 9, our algorithm properly places the major characters in more than one communities. Valjean, a protagonist in *Les Misérables,* and his nemesis (i.e., the police officer Javert), who connect to about 48% of all characters, are assigned to several communities composed of their respective adherents. In addition, our algorithm also classifies other major characters of the novel into their proper communities. As an example, consider *Gavroche*. He is assigned to two communities: the red and the purple communities. The links connecting Gavroche, his family members and their friends are organized in the purple community, and the links connecting him with people associated with the battle are classified into the red community.

*3.2.3. Word associations*

To further evaluate the efficacy of UELC, we considered next a word association network as our third example of real-world networks. The network was constructed from the South Florida Free Association norm list [17]. The data can be represented by a directed, weighted network with words as nodes, their relationships as links and links quantifying the degrees of association between pairs of words. Following the practice in [5], we reduced this network to an undirected and unweighted network by ignoring link directions and weights, and removed the links with weights less than $w^*=0.025$. The resulting network contains 5017 nodes and 29148 links, which is substantially larger and more complex than the two real networks analyzed earlier. To accommodate our analysis, we adopt the way used by Palla et al. [5] to evaluate our partition result.

Following the method in [5], we investigated four statistical properties to characterize the community structures of the word association network. These properties include the community size $s^{com}$ (i.e., the number of nodes in a community), the overlap size $s^{ov}$ (i.e., the number of nodes two communities share), the membership number $m$ (i.e., the number of communities that a node belongs to) and the community degree $d^{com}$ (i.e., the number of communities that have common nodes with a particular community). Figure 10 shows the four distributions characterizing the global community structure revealed by UELC. For comparison, it also shows the statistics for the Clique Percolation Method (CPM) [5] which is the most prominent algorithm for finding overlapping communities. The red triangles correspond to the result of UELC, and the blue asterisks refer to the result of CPM. As shown in Figure 10(a), the behavior of the cumulative community size distribution $P(s^{com})$ for UELC is close to a power law, and incredibly similar to the behavior of CPM. This phenomenon, to some extent, indicates the community structure found by UELC is sound. The cumulative overlap size distributions $P(s^{ov})$ for both UELC and CPM, shown in Figure 10(b), follow approximately power laws, although the ranges of overlap sizes are different for UELC and CPM. Figure 10(c) displays the cumulative distributions of the membership numbers $P(m)$ for UELC and CPM, which also follow closely power law distributions. The membership number for UELC is in general greater than that for CPM, indicating that the former can capture more overlap. Finally, as shown in Figure 10(d), the cumulative distributions of the community degree $P(d^{com})$ for both UELC and CPM follow approximately exponential distributions. The $d^{com}$ for UELC is larger than that for CPM over nearly the entire range of community degree, indicating that UELC is able to extract more highly overlapping communities.

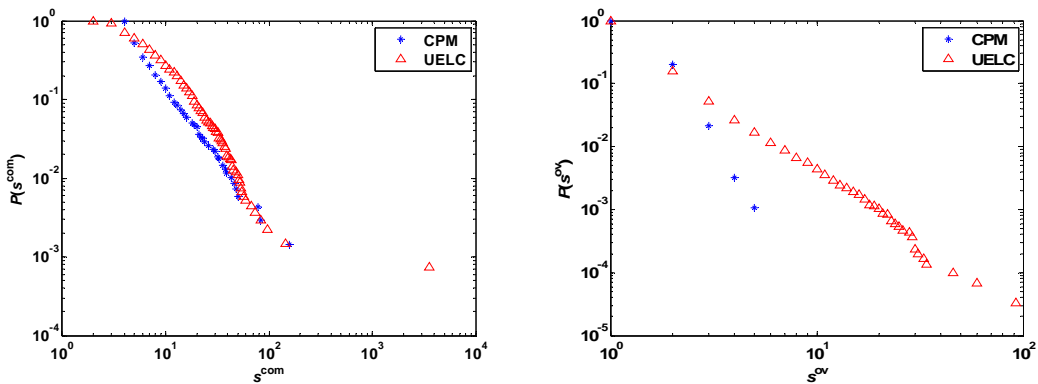

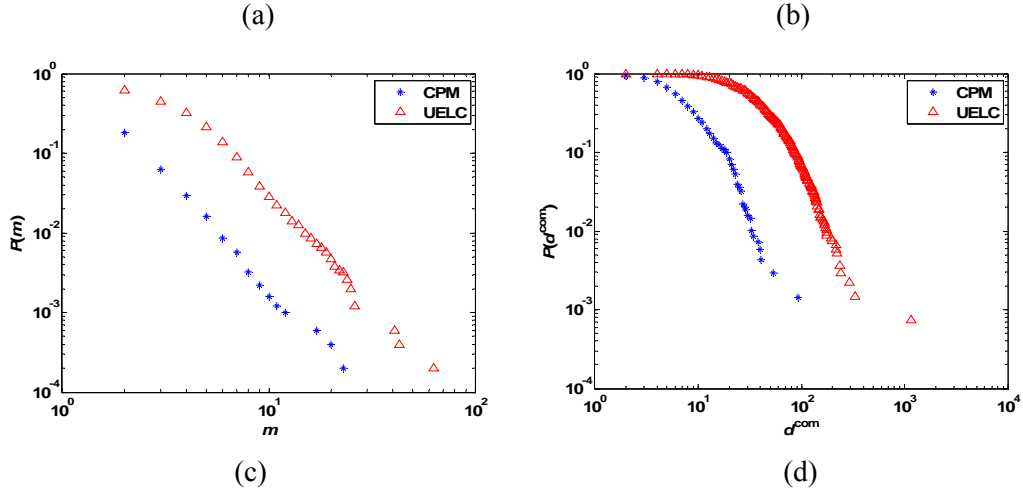

(c)                                              (d)

**Figure 10.** Results comparing UELC and CPM on a word association network, measured in four statistics: the cumulative distributions for (a) community size; (b) overlap size; (c) membership number; and (d) community degree. A statistic for clique percolation is also included for comparison.

## 4. Extend UELC to finding node communities

Although the UELC method was developed for discovering communities of links in networks, overlapping communities of nodes can be naturally derived after a link partition is obtained. Furthermore, it can also be readily extended to detecting nonoverlapping communities of nodes, which is the topic of this section.

Recall the stochastic process of link-node-link random walk. In its formulation, $\alpha^l(e_{ij})$ in (2) denotes the probability that the agent starts from a source position and ultimately arrives at a destination link $e_{ij}$ with $l$ steps. We can use $\alpha^l$ to compute the probability that the agent arrives at an arbitrary node $i$ on the next step from the current link state. In particular, this probability can be computed by,

$$\psi^l(i) = \frac{1}{2}\sum_{j \in Neb(i)} \alpha^l(e_{ij}) \qquad (9)$$

where $Neb(i)$ is the neighbor set of node $i$. According to the analysis in Sec. 2.2.2, $\alpha^\infty(e_{ij}) = 1/m$ for each link $e_{ij}$, where $m$ is the number of links of a network. Thus, we can derive the stable state of $\psi^\infty(i)$ on the basis of $\alpha^\infty(e_{ij})$ by,

$$\psi^\infty(i) = d_i \cdot (\frac{1}{2} \cdot \frac{1}{m}) = \frac{d_i}{\sum_r d_r} \qquad (10)$$

where $d_i$ is the degree of node $i$.

Similarly to the original UELC for finding link communities, to obtain a bipartition node structure of a network, we can take $\psi^\infty$ as a cutoff vector to divide the associated probabilities into two parts. The nodes with associated probabilities greater or less than their stable probabilities form, respectively, two separate communities. Further, in order to improve the performance of the new method, a simple modification can be introduced to the bipartition process: we move every node that is not in the same community with most of its neighbors to the other community. This step, which is easy to implement and cost little computationally comparing with the calculation of the aggregated transition probability distribution, improves the quality of our results obviously.

The stopping criterion of the recursive bisection process needs to be revised accordingly. When one part of the local node bipartition has a lower density than that of the current network (or subnetwork), the recursive process terminates.

We experimentally evaluated the performance of the extended UELC for finding node communities in the LFR benchmark networks [25]. These networks are characterized by heterogeneous distributions of node degrees and community sizes, which are challenging for community detection algorithms. Again, Newman's

EM algorithm was selected for comparison. Furthermore, we employed the widely used accuracy measure of Normalized Mutual Information (NMI) [26] as the quality metric, which has been regarded as a comparative fair metric compared with the other measures [26].

Following the experiment design of [27], we set the parameters for the LFR benchmark networks as follows. The network size $n$ was set to 1000 or 5000, the minimum community size $c_{min}$ was set to 10 or 20, and the mixing parameter $\mu$ (each node shares a fraction $\mu$ of its edges with nodes in other communities) was varied from 0 to 0.8 with an increment of 0.05. We kept the remaining parameters constant: the average degree $d$ is 20, the maximum degree $d_{max}$ is 2.5×$d$, the maximum community size $c_{max}$ is 5×$c_{min}$, and the exponents of the power-law distribution of node degrees $\tau_1$ and community sizes $\tau_2$ are -2 and -1, respectively.

Figure 11 shows the NMI accuracies of the two algorithms compared as a function of the mixing parameter $\mu$. As shown, in most range of mixing parameter $\mu$, such as 0<$\mu$<0.6, the accuracy of UELC is higher than that of Newman's EM. In the remaining range, such as 0.6≤$\mu$≤0.8, the accuracy of UELC is also very close to that of the EM algorithm. Note that we fixed and provided the actual community number to the EM algorithm, while the UELC algorithm was not given such *a priori* information, so that the comparison was in favor of the former.

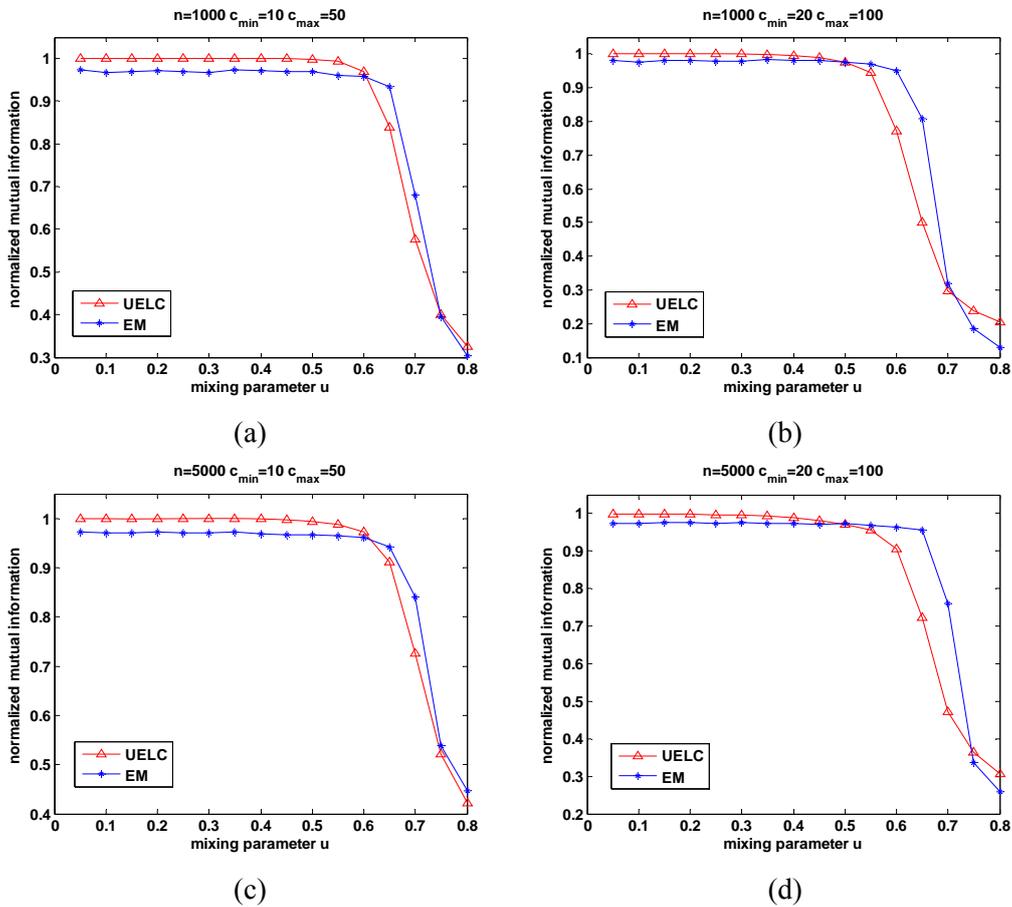

**Figure 11**. Comparison of UELC and Newman's EM algorithms in terms of NMI accuracy on the LFR benchmark networks. Each data point in the figure is an average over 50 network instances. (a) Results on small networks with small communities ($n$ = 1000, $c_{min}$ = 10, $c_{max}$ = 50). (b) Results on small networks with big communities ($n$ = 1000, $c_{min}$ = 20, $c_{max}$ = 100). (c) Results on big networks with small communities ($n$ = 5000, $c_{min}$ = 10, $c_{max}$ = 50). (d) Results on big networks with big communities ($n$ = 5000, $c_{min}$ = 20, $c_{max}$ = 100).

## 5. Discussion and conclusion

We have developed a link dynamics based algorithm, namely UELC, for the identification of link communities in complex networks. The key idea behind UELC rests on the connection between the community structures of networks and some properties of link dynamics process of a random walk, based on which we were able to divide the edges of a network into two separable parts. Specifically, the new algorithm runs in a iterative and recursive procedure of two phases: firstly, the ULC subroutine reveals a bipartition structure of the network by evaluating the (aggregated) transition probability distribution of a stochastic process of link-node-link random walk; next, the two link modules emerged by ULC are extracted by the ELC subroutine, which makes use of the local mixing properties hidden in the Markov chain of the above stochastic process. An iterative procedure is introduced to recursively subdivide the network structures, giving rise to the integrated UELC algorithm. The algorithm was shown to be highly effective in discovering link community structures of synthetic benchmarks and some real-world networks. UELC can also be extended to effectively extracting communities of nodes on some real-world networks.

It is noteworthy to mention a similar previous work done by Evans et al. [4]. Both Evans's and our UELC methods make use of a link-node-link random walk. However, the two methods are also significantly different. Evans's method first transforms a targeted network into a weighted line graph, which is based on the link-node-link random walk. It then applies an existing algorithm for node partition on this weighted line graph. In particular, this method adopts the fast modularity optimization method proposed by Blondel et al. [28]. Different from Evans's method, UELC is a link dynamics based algorithm. It first employs a stochastic process of link-node-link random walk to unfold a bipartition structure of a network. It then utilizes the local mixing properties hidden in the Markov chain of the stochastic process to extract two emerged link communities. This bipartition method can be further extended to a complete algorithm by introducing an iterative subdivision strategy, making the algorithm suitable for discovering hierarchical link partitions of a network.

Our current work can be improved and extended in multiple directions. In the current work, we mainly focused on a technique that divides a network into two link communities, which is based on transition probability distributions of random walks. Furthermore, we designed a stopping criterion by exploiting an intrinsic property of link densities. This termination condition cooperates well with the bipartition technique. Nevertheless, this termination criterion may potentially be improved. For example, the current termination criterion may produce one good link partition, but does not provide an option to deal with different partitioning resolutions, which may be meaningful for a real-world problem instance. Thus, in the future, we intend to further study the relations between the Markov bipartition method and its termination condition, aiming at designing a more suitable strategy to control the successive bipartition process and to accommodate network analysis with a desirable resolution. Another area for further extension is to consider node and link communities at the same time. A real-world network can contain not only node communities but also link communities. Our final UELC method, as we discussed in detail in the previous sections, can be applied to identify link and node communities separately. It is unable to identify both these two patterns at the same time. In our future work, we also plan to develop an integrative to address this issue.

**Acknowledgments**
We thank the anonymous reviewers for their constructive comments and suggestions and Brian Ball for providing some programs used in this study. This work was supported by National Natural Science Foundation of China (60973088, 61133011, 60873149), Scholarship Award for Excellent Doctoral Student granted by Ministry of Education of China (450060454018), Program for New Century Excellent Talents in University of China (NCET-11-0204), and National Science Foundation of US (DBI-0743797).

**Appendix A: framework of the UELC algorithm**
    **Algorithm** $C$ = UELC($N$) //$N$ is a network, $C$ is a link partition of $N$
  1. $C = \{E(N)\}$; // $E(N)$ denotes the edge set of $N$
  2. Unfold a bipartition structure of links for network $N$;
  // by ULC which is described in Sec. 2.2.1
  3. Extract the two emerged link modules $N_1$ and $N_2$;
  // by ELC which is described in Sec. 2.2.2

// $E(N_1) \cap E(N_2) = \Phi$, $E(N_1) \cup E(N_2) = E(N)$
4. If this bipartition result is not acceptable, return $C$;
// the judging condition is introduced in Sec. 2.3
5. $C_1$ = UELC($N_1$);
6. $C_2$ = UELC($N_2$);
7. Return $C = C_1 \cup C_2$.

## References

[1] Girvan M and Newman M E J, *Community structure in social and biological networks*, 2002 *Proc. Natl. Acad. Sci.* **9** 7821

[2] Fortunato S, *Community detection in graphs*, 2010 *Phys. Rep.* **486** 75

[3] Ahn Y, Bagrow J P, and Lehmann S, *Link communities reveal multiscale complexity in networks*, 2010 *Nature* **466** 761

[4] Evans T S and Lambiotte R, *Line graphs, link partitions, and overlapping communities*, 2009 *Phys. Rev.* E **80** 016105

[5] Palla G, Derenyi I, Farkas I and Vicsek T, *Uncovering the overlapping community structures of complex networks in nature and society*, 2005 *Nature* **435** 814

[6] Evans T S and Lambiotte R, *Line graphs of weighted networks for overlapping communitie,* 2010 *Eur. Phys. J.* B **77** 265

[7] Kim Y and Jeong H, *Map equation for link communities*, 2011 *Phys. Rev.* E **84** 026110

[8] Rosvall M and Bergstrom C T, *Maps of random walks on complex networks reveal community structure*, 2008 *Proc. Natl. Acad. Sci.* **105** 1118

[9] Ball B, Karrer B, and Newman M E J, *Efficient and principled method for detecting communities in networks*, 2011 *Phys. Rev.* E **84** 036103

[10] Pan L, Wang C, Xie J, and Liu M, *Detecting link communities based on local approach,* 2011 *ICTAI'11: Proc. 23rd IEEE Int. Conf. on Tools with Artificial Intelligence* (Boca Raton, Florida, USA: IEEE) pp 884-86

[11] Shannon P, Markiel A, Ozier O, Baliga N S, Wang J T, Ramage D, Amin N, Schwikowski B and Ideker T, *Cytoscape: a software environment for integrated models of biomolecular interaction networks*, 2003 *Genome Res.* **13** 2498

[12] Eades P, *A heuristic for graph drawing*, 1984 *Congr. Numer.* **42** 142

[13] Yang B, Liu J and Feng J, *On the spectral characterization and scalable mining of network communities*, 2012 *IEEE Trans. Knowl. Data Eng.* **24** 326

[14] Golub G H and Loan C F V, *Matrix Computation,* 1989 Johns Hopkins Univ. Press

[15] Zachary W W, *An information flow model for conflict and fission in small groups,* 1977 *J. Anthropol. Res.* **33** 452

[16] Knuth D E, *The Stanford GraphBase: A Platform for Combinatorial Computing*, 1993 Addison–Wesley, Reading, MA

[17] Nelson D L, McEvoy C L, and Schreiber T A, The University of South Florida, word association, rhyme, and word fragment norms, http://w3.usf.edu/FreeAssociation/

[18] Newman M E J and Girvan M, *Finding and evaluating community structure in networks,* 2004 *Phys. Rev.* E **69** 026113

[19] Newman M E J, *Modularity and community structure in networks,* 2006 *Proc. Natl. Acad. Sci. USA* **103** 8577

[20] Reichardt J, Bornholdt S, *Statistical mechanics of community detection*, 2006 *Phys. Rev.* E **74** 016110

[21] Fortunato S and Barthélemy M, *Resolution limit in community detection,* 2007 *Proc. Natl. Acad. Sci. U.S.A.* **104** 36

[22] Lancichinetti A and Fortunato S, *Limits of modularity maximization in community detection*, 2011 *Phys. Rev.* E **84** 066122

[23] Good B H, Montjoye de Y-A and Clauset A, *The performance of modularity maximization in practical contexts*, 2010 *Phys. Rev.* E **81** 046106

[24] The software used in this paper can be found in ftp://jindijlu:dd@59.72.0.62/

[25] Lancichinetti A, Fortunato S and Radicchi F, *Benchmark graphs for testing community detection algorithms,* 2008 *Phys. Rev.* E **78** 046110

[26] Danon L, Duch J, Diaz-Guilera A and Arenas A, *Comparing community structure identification,* 2005 *J. Stat. Mech.* **2005** P09008

[27] Lancichinetti A and Fortunato S, *Community detection algorithms: A comparative analysis,* 2009 *Phys. Rev.* E **80** 056117

[28] Blondel V D, Guillaume J L, Lambiotte R, Lefebvre E, *Fast unfolding of communities in large networks,* 2008 *J. Stat. Mech: Theory Exp.* P10008